\documentstyle[12pt,fleqn]{article}
\def\beq{\begin{equation}}
\def\eeq{\end{equation}}

\begin{document}

\vspace*{15mm}
{\Large {\bf Quantum Entanglement:\\[2mm] 
Criteria and Collective Tests}} \vfill

Asher Peres$^*$ \\[5mm]
Department of Physics, Technion---Israel Institute of Technology,
32\,000 Haifa, Israel\vfill

{\bf Abstract}\\[5mm]
The state of a quantum system, consisting of two distinct subsystems,
is called {\it separable\/} if it can be prepared by two distant
experimenters who receive instructions from a common source, via
classical communication channels. A necessary condition is derived and
is shown to be more sensitive than Bell's inequality for detecting
quantum inseparability.  Moreover, collective tests of Bell's inequality
(namely, tests that involve several composite systems simultaneously)
may sometimes lead to a violation of Bell's inequality, even if the
latter is satisfied when each composite system is tested separately.
\vfill

\vfill

\noindent$^*$\,Electronic address: peres@photon.technion.ac.il\vfill

\newpage
\parindent 5mm

\noindent{\bf 1. Introduction}\bigskip

\noindent Quantum entanglement is a physical resource. Applications
include secure communication between distant observers~[1], faithful
tele\-portation of an unknown quantum state to an unknown location~[2],
quantum computation, and in particular error correction codes~[3].
Quantum entanglement is subtler than  classical correlation. A composite
quantum system can be prepared in a prescribed state $\rho$, with
distant correlations, by two observers who only use local raw materials
and receive instructions from a common source by a classical
communication channel.  Such a preparation method yields a density
matrix which is {\it separable\/} into a sum of direct products,

\beq \rho=\sum_K w_K\,\rho_K'\otimes\rho_K'', \label{sep}\eeq
where the positive weights $w_K$ satisfy $\sum w_K=1$, and where
$\rho_K'$ and $\rho_K''$ are density matrices for the two subsystems.  A
separable system represented by the above $\rho$ is {\it correlated\/},
but it is not {\it entangled\/}: it always satisfies Bell's
inequality~[4].

That inequality was originally derived for solving a completely
different problem: is quantum theory compatible with an underlying
pseudo-classical ``subquantum'' background (deterministic or possibly
stochastic)? Such a post-quantum theory would presumably involve
additional ``hidden'' variables, and the statistical predictions of
ordinary quantum theory would be reproduced by performing suitable
averages over these hidden variables. Bell~[4] was the first to show
that if the constraint of {\it locality\/} is imposed on the hidden
variables (namely, if the hidden variables of two distant quantum
systems are themselves separable into two distinct subsets), then
there is an upper bound to the correlations of results of measurements
that can be performed on the two distant systems. That bound is violated
by some states in quantum mechanics, for example the singlet state of
two \mbox{spin-$1\over2$} particles. It is not quantum theory itself
which is nonlocal. It is any pseudo-classical theory which attempts to
mimic quantum effects that is necessarily nonlocal and contextual.

A variant of Bell's inequality, more general and more useful for
experimental tests, was derived by Clauser, Horne, Shimony, and
Holt (CHSH)~[5]. It can be written

\beq |\langle{AB}\rangle+\langle{AB'}\rangle+\langle{A'B}\rangle
  -\langle{A'B'}\rangle|\leq 2. \label{CHSH}\eeq
On the left hand side, $A$ and $A'$ are two operators that can be
measured by an observer, conventionally called Alice. These
operators do not commute (so that Alice has to choose whether to
measure $A$ or $A'$) and each one is normalized to unit norm (the norm 
of an operator is defined as the largest absolute value of any of its
eigenvalues). Likewise, $B$ and $B'$ are two normalized noncommuting
operators, any one of which can be measured by another, distant
observer (Bob). Note that each one of the {\it expectation\/} values in
Eq.~(\ref{CHSH}) can be calculated by means of quantum theory, if the
quantum state is known, and it is also experimentally observable, by
repeating the measurements sufficiently many times, starting each time
with identically prepared pairs of quantum systems. The validity of
the CHSH inequality, for {\it all\/} combinations of
measurements independently performed on both systems, is a necessary 
condition for the possible existence of a local hidden variable (LHV)
model for the results of these measurements. It is not in general a
sufficient condition, as will be shown below.

Note that, in order to test Bell's inequality, the two distant observers
independently {\it measure\/} subsytems of a composite quantum system, 
and then they {\it report\/} their results to a common site where that
information is analyzed~[6]. This is the converse of the situation that
was mentioned above, where the two observers {\it received\/}
instructions from a common center. There are density matrices for which
it can be proved that no such set of instructions exists, and yet Bell's
inequality is satisfied~[7--10].  I shall derive below a simple
algebraic test which is a necessary condition for the existence of the
decomposition~(\ref{sep}). I shall then give some examples showing that
this new criterion is more restrictive than Bell's inequality, or than
the $\alpha$-entropy inequality~[11].\\[7mm]

\noindent{\bf 2. Separability of density matrices}\bigskip

\noindent The derivation of the separability condition is easiest when
the density matrix elements are written explicitly, with all their
indices~[6]. For example, Eq.~(\ref{sep}) becomes
\beq \rho_{m\mu,n\nu}=
  \sum_K w_K\,(\rho'_K)_{mn}\,(\rho''_K)_{\mu\nu}. \eeq
Latin indices refer to the first subsystem, Greek indices to the second
one (the sub\-systems may have different dimensions). Note that this
equation can always be satisfied if we replace the quantum density
matrices by classical Liouville functions (and the discrete indices are
replaced by canonical variables, {\bf p} and {\bf q}). The reason is
that the only constraint that a Liouville function has to satisfy is
being non-negative. On the other hand, we want quantum density matrices
to have non-negative {\it eigenvalues\/}, rather than non-negative
elements, and the latter condition is more difficult to satisfy.

It is noteworthy that any set of instructions that Alice and Bob may
receive is inherently ambiguous, if these observers have no way of
ascertaining that what one of them calls $i=\sqrt{-1}$ is the same as
$i$ (not $-i$) of the other observer. Indeed, the only way of explaining
to someone what $i$ and $-i$ actually are is by referring to a material
(chiral) object, such as a screw.  If we are restricted to the use of
verbal instructions for preparing $\rho$, the result may be either
$\rho$ or $\rho^*=\rho^T$. Therefore Alice and Bob may also end up with
the ``dual'' matrix

\beq \sigma=\sum_A w_A\,(\rho_A')^T\otimes\rho_A''. \label{sig}\eeq

This ambiguity is an important clue for solving the following problem.
Let all the matrix elements of $\rho$ be given. Is a decomposition such
as in Eq.~(\ref{sep}) possible?  A simple condition is readily obtained
by defining a new matrix,

\beq \sigma_{m\mu,n\nu}\equiv\rho_{n\mu,m\nu}. \eeq
The Latin indices of $\rho$ have been transposed, but not the Greek
ones. This is not a unitary transformation but, nevertheless, the
$\sigma$ matrix is Hermitian. Whenever Eq.~(\ref{sep}) is valid,
the transposed matrices $(\rho'_A)^T\equiv(\rho'_A)^*$ are
non-negative matrices with unit trace, so that they also are legitimate
density matrices. It follows that {\it none of the eigenvalues of
$\sigma$ is negative\/}. This is a necessary condition for
Eq.~(\ref{sep}) to hold~[12].

Note that the eigenvalues of $\sigma$ are invariant under separate
unitary transformations, $U'$ and $U''$, of the bases used by the two
observers. In such a case, $\rho$ transforms as
\beq \rho\to (U'\otimes U'')\,\rho\,(U'\otimes U'')^\dagger, \eeq
and we then have
\beq \sigma\to
 (U'^T\otimes U'')\,\sigma\,(U'^T\otimes U'')^\dagger, \eeq
which also is a unitary transformation, leaving the eigenvalues of
$\sigma$ invariant.

As an example, consider a pair of spin-$1\over2$ particles in 
an impure singlet state, consisting of a singlet fraction $x$ and
a random fraction $(1-x)$~[13]. Note that the ``random
fraction'' $(1-x)$ also includes singlets, mixed in equal proportions
with the three triplet components. We have
\beq \rho_{m\mu,n\nu}=x\,S_{m\mu,n\nu}+
  (1-x)\,\delta_{mn}\,\delta_{\mu\nu}\,/4, \label{x}\eeq
where the density matrix for a pure singlet is given by
\beq S_{01,01}=S_{10,10}=-S_{01,10}=-S_{10,01}=\mbox{$1\over2$},
 \label{singlet} \eeq
and all the other components of $S$ vanish. (The indices 0 and 1 refer
to any two ortho\-gonal states, such as ``up'' and ``down.'') A
straightforward calculation shows that $\sigma$ has three eigenvalues
equal to $(1+x)/4$, and the fourth eigenvalue is $(1-3x)/4$. This lowest
eigenvalue is positive if $x<{1\over3}$, and the separability criterion
is then fulfilled. This result may be compared with other criteria:
Bell's inequality holds for $x<1/\sqrt{2}$, and the $\alpha$-entropic
inequality~[11] for $x<1/\sqrt{3}$. These are therefore much
weaker tests for detecting inseparability than the condition that was
derived here.

In this particular case, it happens that this necessary condition is
also a sufficient one. It is indeed known that if $x<{1\over3}$ it is
possible to write $\rho$ as a mixture of unentangled product
states~[14].  This suggests that the necessary condition derived above
($\sigma$ has no negative eigenvalue) might also be sufficient for any
$\rho$. A proof of this conjecture was indeed recently obtained~[15] for
composite systems having dimensions $2\times2$ and $2\times3$. However,
for higher dimensions, the present necessary condition is {\it not\/} a
sufficient one. Some counter\-examples have been constructed, with
dimensions $2\times4$ and $3\times3$~[16]. It was also shown~[16] that
the decomposition of any separable $\rho$ requires at most 
$({\rm dim}~{\cal H})^2$ terms of rank one. This property is an
important step toward finding an {\it efficient\/} algorithm for the
decomposition of any separable $\rho$. Such an algorithm, which could
also indicate that $\rho$ is not separable, is still lacking at the time
of writing.

As a second example, consider a mixed state consisting of a
fraction $x$ of the pure state $a|01\rangle+b|10\rangle$ (with
$|a|^2+|b|^2=1$), and fractions $(1-x)/2$ of the pure states
$|00\rangle$ and $|11\rangle$. We have

\beq \rho_{00,00}=\rho_{11,11}=(1-x)/2,\eeq
\beq \rho_{01,01}=x|a|^2, \eeq
\beq \rho_{10,10}=x|b|^2, \eeq
\beq \rho_{01,10}=\rho_{10,01}^*=xab^*, \eeq
and the other elements of $\rho$ vanish. It is easily seen that
the $\sigma$ matrix
has a negative determinant, and thus a negative eigenvalue, when 
\beq x>(1+2|ab|)^{-1}. \eeq
This is a lower limit than the one for a violation of Bell's inequality,
which requires~[10]
\beq x>[1+2|ab|(\sqrt{2}-1)]^{-1}. \eeq

An even more striking example is the mixture of a singlet and a
maximally polarized pair:
\beq \rho_{m\mu,n\nu}=x\,S_{m\mu,n\nu}+
  (1-x)\,\delta_{m0}\,\delta_{n0}\,\delta_{\mu0}\,\delta_{\nu0}.\eeq
For any positive $x$, however small, this state is inseparable, because
$\sigma$ has a negative eigenvalue $(-x/2)$. On the other hand, the
Horodecki criterion~[17] gives a very generous domain to the
validity of Bell's inequality: $x\leq 0.8$.\\[7mm]

\noindent{\bf 3. Collective tests for nonlocality}\bigskip

\noindent The weakness of Bell's inequality as a test for inseparability
is due to the fact that the only use made of the density matrix
$\rho$ is for computing the probabilities of the various outcomes of
tests that may be performed on the subsystems of a {\it single\/}
composite system. On the other hand, an experimental verification of
that inequality necessitates the use of {\it many\/} composite systems,
all prepared in the same way.  However, if many such systems are
actually available, we may also test them collectively, for example two
by two, or three by three, etc., rather than one by one. If we do that,
we must use, instead of $\rho$ (the density matrix of a single system),
a {\it new\/} density matrix, which is $\rho\otimes\rho$, or
$\rho\otimes\rho\otimes\rho$, in a higher dimensional space. It will now
be shown that there are some density matrices $\rho$ that satisfy
Bell's inequality, but for which $\rho\otimes\rho$, or
$\rho\otimes\rho\otimes\rho$, etc., violate that inequality~[18].

The example that will be discussed is that of the Werner
states~[7] defined by Eq.~(\ref{x}). Let us consider
$n$ Werner pairs. Each one of the two observers has $n$
particles (one from each pair). They proceed as follows. First, they
subject their $n$-particle systems to suitably chosen local unitary
transformations, $U$, for Alice, and $V$, for Bob. Then, they test
whether each one of the particles labelled 2, 3, \ldots, $n$, has spin
up (for simplicity, it is assumed that all the particles are
distinguishable, and can be labelled unambiguously). Note that any
other test that they can perform is unitarily equivalent to the one for
spins up, as this involves only a redefinition of the matrices $U$ and
$V$. If any one of the $2(n-1)$ particles tested by Alice and Bob shows
spin down, the experiment is considered to have failed, and the two
observers must start again with $n$ new Werner pairs.

A similar elimination of ``bad'' samples is also inherent to any
experimental procedure where a failure of one of the detectors to fire
is handled by discarding the results registered by all the other
detectors:  only when {\it all\/} the detectors fire are their results
included in the statistics. This obviously requires an exchange of {\it
classical\/} information between the observers. (There is a controversy
on whether a violation of Bell's inequality with post\-selected
data~[19] is a valid test for non\-locality~[20]. I shall not discuss
this issue here; I only examine whether or not Bell's inequality is
violated by the post\-selected data.)

The calculations shown below will refer to the case $n=3$, for
definiteness.  The generalization to any other value of $n$ is
straightforward. Spinor indices, for a single \mbox{spin-$1\over2$}
particle, will take the values 0 (for the ``up'' component of spin) and
1 (for the ``down'' component). The 16 components of the density matrix
of a Werner pair, consisting of a singlet fraction $x$ and a random
fraction $(1-x)$, are, in the standard direct product basis:
\beq \rho_{mn,st}=x\,S_{mn,st}+(1-x)\,\delta_{ms}\,\delta_{nt}\,/4,\eeq
where I am now using only Latin indices, contrary to what I did in
Eq.~(\ref{x}); this is because Greek indices will be needed for another
purpose, as will be seen soon. Thus, now, the indices $m$ and $s$ refer
to Alice's particle, and $n$ and $t$ to Bob's particle.

When there are three Werner pairs, their combined density matrix is a
direct product $\rho\otimes\rho'\otimes\rho''$, or explicitly,
$\rho_{mn,st}\,\rho_{m'n',s't'}\,\rho_{m''n'',s''t''}$. The result of
the unitary transformations $U$ and $V$ is
\beq \rho\otimes\rho'\otimes\rho''\to
 (U\otimes V)\,(\rho\otimes\rho'\otimes\rho'')\,
 (U^\dagger\otimes V^\dagger). \label{newrho} \eeq
Explicitly, with all its indices, the $U$ matrix satisfies the unitarity
relation
\beq \sum_{mm'm''}U_{\mu\mu'\mu'',mm'm''}\;
 U^*_{\lambda\lambda'\lambda'',mm'm''}=
 \delta_{\mu\lambda}\;\delta_{\mu'\lambda'}\;\delta_{\mu''\lambda''}.
 \label{unitary}\eeq
In order to avoid any possible ambiguity, Greek indices (whose values
are also 0 and 1) are now used to label spinor components {\it after\/}
the unitary transformations. Note that the indices without primes refer
to the two particles of the first Werner pair (the only ones that are
not tested for spin up) and the primed indices refer to all the other
particles (that are tested for spin up).  The $V_{\nu\nu'\nu'',nn'n''}$
matrix elements of Bob's unitary transformation satisfy a relationship
similar to (\theequation). The generalization to a larger number of
Werner pairs is obvious.

After the execution of the unitary transformation (\ref{newrho}), Alice
and Bob have to test that all the particles, except those labelled by
the first (unprimed) indices, have their spin up. They discard any set
of $n$ Werner pairs where that test fails, even once. The density matrix
for the remaining ``successful'' cases is thus obtained by retaining, on
the right hand side of Eq.~(\ref{newrho}), only the terms whose primed
components are zeros, and then renormalizing the resulting matrix to
unit trace. This means that only two of the $2^n$ rows of the $U$
matrix, namely those with indices 000\ldots\ and 100\ldots, are relevant
(and likewise for the $V$ matrix).  The elimination of all the other
rows greatly simplifies the problem of optimizing these matrices. We
shall thus write, for brevity,
\beq U_{\mu 00,mm'm''}\to U_{\mu,mm'm''}, \eeq
where $\mu=0,1$. Then, on the left hand side of Eq.~(\ref{unitary}), we
effectively have two unknown row vectors, $U_0$ and $U_1$, each one with
$2^n$ components (labelled by Latin indices  $mm'm''$). These vectors
have unit norm and are mutually orthogonal. Likewise, Bob has two
vectors, $V_0$ and $V_1$.  The problem is to optimize these four vectors
so as to make the expectation value of the Bell operator~[21],
\beq C:=AB+AB'+A'B-A'B', \eeq
as large as possible.

The optimization proceeds as follows. The new density matrix, for the
pairs of spin-\mbox{$1\over2$} particles that were {\it not\/} tested
by Alice and Bob for spin up
(that is, for the first pair in each set of $n$ pairs), is

\beq (\rho_{\rm new})_{\mu\nu,\sigma\tau}=
 N\, U_{\mu,mm'm''}\,V_{\nu,nn'n''}\;\rho_{mn,st}\;\rho_{m'n',s't'}\;
 \rho_{m''n'',s''t''}\,U^*_{\sigma,ss's''}\,V^*_{\tau,tt't''},\eeq
where $N$ is a normalization constant, needed to obtain unit trace
($N^{-1}$  is the probability that all the ``spin up'' tests were
successful). We then have~[17], for fixed $\rho_{\rm new}$ and
all possible choices of $C$,

\beq \max\,[{\rm Tr}\,(C\rho_{\rm new})]=2\sqrt{M}, \label{M}\eeq
where $M$ is the sum of the two largest eigenvalues of the real
symmetric matrix $T^\dagger T$, defined by

\beq T_{pq}:={\rm Tr}\,[(\sigma_p\otimes\sigma_q)\,\rho_{\rm new}].
 \label{T} \eeq
(In the last equation, $\sigma_p$ and $\sigma_q$ are the Pauli spin
matrices.) Our problem is to find the vectors $U_\mu$ and $V_\nu$ that
maximize $M$.

At this point, some simplifying assumptions are helpful.  Since all
matrix elements $\rho_{mn,st}$ are real, it is reasonable to restrict
the search to vectors $U_\mu$ and $V_\nu$ that have only real
components.  Furthermore, the situations seen by Alice and Bob are
completely symmetric, except for the opposite signs in the standard
expression for the singlet state:

\beq \textstyle{\psi=
 \left[{1\choose0}{0\choose1}-{0\choose1}{1\choose0}\right]
   \;/\sqrt{2}.} \eeq
These signs can be made to become the same by redefining the
basis, for example by representing the ``down'' state of Bob's particle
by the symbol ${0\choose-1}$, {\it without\/} changing the basis used
for Alice's particle. This unilateral change of basis is equivalent a
substitution
\beq V_{\nu,nn'n''}\to(-1)^{\nu+n+n'+n''}\,V_{\nu,nn'n''},\eeq
on Bob's side. The minus signs in Eq.~(\ref{singlet}) also disappear,
and there is complete symmetry for the two observers. It is then
plausible that, with the new basis, the optimal $U_\nu$ and $V_\nu$
are the same. Therefore, when we return to the original basis and
notations, they satisfy
\beq V_{\nu,nn'n''}=(-1)^{\nu+n+n'+n''}\,U_{\nu,nn'n''}.\eeq
We shall henceforth restrict our search to pairs of vectors that satisfy
this relation.

After all the above simplifications, the problem that has to be solved
is the following: find two mutually orthogonal unit vectors, $U_0$ and
$U_1$, each one with $2^n$ real components, that maximize the value of
$M(U)$ defined by Eqs.~(\ref{M}) and~(\ref{T}). This is a standard
optimization problem which can be solved numerically.  Since the
function $M(U)$ is bounded, it has at least one maximum.  It may,
however, have more than one: there may be several distinct local
maxima with different values. A numerical search leads to one
of these maxima, but not necessarily to the largest one. The outcome
may depend on the initial point of the search. It is therefore
imperative to start from numerous randomly chosen points in order to
ascertain, with reasonable confidence, that the largest maximum has
indeed been found.\\[7mm]

\noindent{\bf 4. Numerical results}\bigskip

\noindent In all the cases that were examined, $M(U)$ turned out to
have a local maximum for the following simple choice:
\beq U_{0,00\ldots}=U_{1,11\ldots}=1, \label{xor}\eeq
and all the other components of $U_0$ and $U_1$ vanish. Recall that the
``vectors'' $U_0$ and $U_1$ actually are two rows, $U_{000\ldots}$
and $U_{100\ldots}$, of a unitary matrix of order $2^n$ (the
other rows are irrelevant because of the elimination of all the
experiments in which a particle failed the spin-up test). In the case
$n=2$, one of the unitary matrices having the property (\ref{xor}) is a
simple permutation matrix that can be implemented by a ``controlled-{\sc
not}'' quantum gate~[22] known as {\sc xor} (exclusive {\sc or}). For
larger values of $n$, matrices that satisfy Eq.~(\ref{xor}) will also be
called {\sc xor}-transformations.

It was found, by numerical calculations, that {\sc xor}-transformations
always are the optimal ones for $n=2$. They are also optimal for $n=3$
when the singlet fraction $x$ is less than 0.57, and for $n=4$ when
$x<0.52$. For larger values of $x$, more complicated forms of $U_0$ and
$U_1$ give better results. For $n=3$, it was recently found that the
optimum is a ``controlled Hadamard transform''

\beq U_{0,000}=U_{0,111}=U_{1,001}=U_{1,100}=1/\sqrt{2}. \eeq
The corresponding result for higher $n$ is not known. Anyway, the
existence of two different sets of maxima may be seen in Fig.~1: there
are discontinuities in the slopes of the graphs for $n=3$ and~4, that
occur at the values of $x$ where the largest value of
$\langle{C}\rangle$ jumps from one local maximum to another one.

For $n=5$, a complete determination of $U_0$ and $U_1$ requires the
optimization of 64 parameters subject to 3 constraints, more than my
workstation could handle in a reasonable time. I therefore considered
only {\sc xor}-transformations, which are likely to be optimal for $x\,
\mbox{\raisebox{-2pt}{{\scriptsize $\stackrel{\textstyle <}{\sim}$}}}\,
0.5$. In particular, for $x=0.5$ (the value that was used in Werner's
original work~[7]), the result is $\langle C\rangle=2.0087$, and the
CHSH inequality is violated. This violation occurs in spite of the
existence of an explicit LHV model that gives correct results if the
Werner pairs are tested one by one.

These results prompt a new question: can we get stronger {\it
inseparability\/} criteria by considering $\rho\otimes\rho$, or higher
tensor products? It is easily seen that no further progress can be
achieved in this way. If $\rho$ is separable as in Eq.~(\ref{sep}), so
is $\rho\otimes\rho$.  Moreover, the partly transposed matrix
corresponding to $\rho\otimes\rho$ simply is $\sigma\otimes\sigma$, so
that if no eigenvalue of $\sigma$ is negative, then
$\sigma\otimes\sigma$ too has no negative eigenvalue.\clearpage

\noindent{\bf Acknowledgment}\bigskip

\noindent This work was supported by the Gerard Swope Fund, and the Fund
for Encouragement of Research.\\[7mm]

\begin{enumerate}
\item Ekert, A. K., Phys. Rev. Lett. {\bf 67}, 661 (1991).
\item Bennett, C. H, Brassard, G., Cr\'epeau, C., Jozsa, R., Peres, A.,
 and Wootters, W. K., Phys. Rev. Lett. {\bf 70}, 1895 (1993).
\item Knill, E., and Laflamme, R., Phys. Rev. A {\bf 55}, 900 (1997).
\item Bell, J. S., Physics {\bf 1}, 195 (1964).
\item Clauser, J. F., Horne, M. A., Shimony, A., and Holt, R. A.,
Phys. Rev. Lett. {\bf 23}, 880 (1969).
\item Peres, A., ``Quantum Theory: Concepts and Methods''
(Kluwer, Dordrecht, 1993) Chapters 5 and 6.
\item Werner, R. F., Phys. Rev. A {\bf 40}, 4277 (1989).
\item Popescu, S., Phys. Rev. Lett. {\bf 72}, 797 (1994);
{\bf 74}, 2619 (1995).
\item Mermin, N. D., in ``Perspectives on Quantum Reality'' (Edited
by R. K. Clifton) (Kluwer, Dordrecht, 1996) pp.~57--71.
\item Gisin, N., Phys. Lett. A {\bf 210}, 151 (1996).
\item Horodecki, R., Horodecki, P., and Horodecki, M., Phys.
Lett. A {\bf 210}, 377 (1996).
\item Peres, A., Phys. Rev. Lett. {\bf 77}, 1413 (1996).
\item Blank, J. and Exner, P., Acta Univ.
Carolinae, Math. Phys. {\bf 18}, 3 (1977).
\item Bennett, C. H., Brassard, G., Popescu, S., Schumacher, B.,
Smolin, J., and Wootters, W. K., Phys. Rev. Lett. {\bf 76}, 722 (1996);
{\bf 78}, 2031 (1997) (E).
\item Horodecki, M., Horodecki, P., and Horodecki, R.,  Phys. Lett. A
{\bf 223}, 1 (1996).
\item Horodecki, P., ``Separability criterion and inseparable mixed
states with positive partial transposition'' (e-print archive:
quant-ph/9605038).
\item Horodecki, R., Horodecki, P., and Horodecki, M.,  Phys.
Lett. A {\bf 200}, 340 (1995).
\item Peres, A., Phys. Rev. A {\bf 54}, 2685 (1996).
\item Peres, A., in ``Quantum Potentiality,
Entanglement, and Passion-at-a-Distance: Essays for Abner Shimony''
(Edited R. S. Cohen, M.~Horne, and J. Stachel) (Kluwer, Dordrecht, 1997).
\item Santos, E., Phys. Rev. Lett. {\bf 66}, 1388 (1991).
\item Braunstein, S. L., Mann, A., and Revzen, M., Phys. Rev.
Lett. {\bf 68}, 3259 (1992).
\item Barenco, A., Deutsch, D., Ekert, A., and Jozsa, R.,  
Phys. Rev. Lett. {\bf 74}, 4083 (1995).

\end{enumerate}\vfill

\parindent 0mm
{\bf Caption of figure}\bigskip

FIG. 1. \ Maximal expectation value of the Bell operator, versus the
singlet fraction in the Werner state, for collective tests performed on
several Werner pairs (from bottom to top of the figure, 1, 2, 3, and 4
pairs, respectively). The CHSH inequality is violated when
$\langle{C\rangle}>2$.

\end{document}